Invited Survey Paper

# Towards mmWave V2X in 5G and Beyond to Support Automated Driving

Kei SAKAGUCHI[†], *fellow*, Ryuichi FUKATSU[†], *student member,* Tao YU[†], *member,* Eisuke FUKUDA[†], *fellow,*
Kim MAHLER[††], *nonmembers,* Robert HEATH[†††], *member,* Takeo FUJII[††††], *fellow*,
Kazuaki TAKAHASHI[†††††], *member,* Alexey KHORYAEV[††††††], *nonmembers,*
Satoshi NAGATA[†††††††], *member,* Takayuki SHIMIZU[††††††††], *member*

**SUMMARY** Millimeter wave provides high data rates for Vehicle-to-Everything (V2X) communications. This paper motivates millimeter wave to support automated driving and begins by explaining V2X use cases that support automated driving with references to several standardization bodies. The paper gives a classification of existing V2X standards: IEEE802.11p and LTE V2X, along with the status of their commercial deployment. Then, the paper provides a detailed assessment on how millimeter wave V2X enables the use case of cooperative perception. The explanations provide detailed rate calculations for this use case and show that millimeter wave is the only technology able to achieve the requirements. Furthermore, specific challenges related to millimeter wave for V2X are described, including coverage enhancement and beam alignment. The paper concludes with some results from three studies, i.e. IEEE802.11ad (WiGig) based V2X, extension of 5G NR (New Radio) toward mmWave V2X, and prototypes of intelligent street with mmWave V2X.

## 1. Introduction

Automated driving is revolutionizing transportation. Most recent commercial efforts in automated driving have targeted fully autonomous operations. In this way, the vehicles are self-driving using Artificial Intelligence (AI) based on data collected from sensors mounted on the vehicle, along with prior stored 3D map data. Automated driving can be enhanced by real-time vehicular communications with surrounding vehicles and Road Side Units (RSUs). This is due to the fact that sensor data from the ego-vehicle is always limited in its field of vision. Adequate communication can help to overcome this shortcoming by exchanging sensor data with other vehicles. The inclusion of sensor data from other vehicles or RSUs will expand the field of perception, lead to better AI decisions and ultimately to safer automated driving. Thus, Vehicular-to-Vehicular (V2V) and more broadly Vehicular-to-Everything (V2X) communications can improve the safety and traffic efficiency of automated driving.

The first generation of V2V use cases was introduced to assist human drivers by exchanging periodic awareness messages and broadcasting warning messages by using V2V [1][2][3]. These use cases were later extended to support road safety services by utilizing V2X [4][5]. Requirements for the first generation V2X, i.e. periodic broadcast of V2X messages (typically, a message of a few hundred bytes every 100 ms), latency of 100 ms, reliability of 99% (cumulative reliability of 2 transmissions), etc., were designed for human drivers. However, with the birth of automated driving technologies, more advanced use cases of V2X were introduced, i.e. vehicles platooning, extended sensors, advanced driving, and remote driving [6][7][8]. In these emerging use cases, larger messages are exchanged via V2X to support "AI drivers" including raw sensor data, vehicles' intention data, coordination, confirmation of future maneuver, etc. These enhanced use cases have far more demanding requirements, such as data rate of 1000 Mbps, latency of 3 ms, reliability of 99.999% (PR.5.6-004, PR.5.6-007 in [6]) for the extended sensors which allows cooperative perception for AI drivers.

To support the first generation V2X use cases of Intelligent Transport Systems (ITS), communication standards of IEEE 802.11p [9] and Long Term Evolution (LTE)-V2X [10][11] were specified in 2010 and 2016 respectively. IEEE802.11p is the basis of Dedicated Short-Range Communication (DSRC) in U.S., ITS-G5 in Europe, and ITS Connect in Japan. DSRC and ITS-G5 are operated in the 5.9 GHz band, while ITS-Connect is in 760 MHz band in Japan. IEEE802.11p supports a data rate of up to 27 Mbps, which is enough for the first generation V2X use cases. LTE-V2X specified in Release 14 by 3rd Generation Partnership Project (3GPP) is an extension of LTE to support not only uplink (UL) and downlink (DL) communication for V2X but also sidelink communication for V2V at the 5.9 GHz ITS band, supporting peak data rate of 28.8 Mbps. The performances of IEEE802.11p and LTE-V2X are sufficient to assist human drivers, but new communication standards are needed to meet higher requirements of enhanced use cases with AI drivers.

[†]. The authors are with Tokyo Institute of Technology, Tokyo, 152-8552 Japan.
[††]. The author is with New York University, New York, NY, 10003, USA.
[†††]. The author is with The University of Texas at Austin, Austin, TX, 78712, USA.
[††††]. The author is with The University of Electro-Communications, Chofu-shi, 182-8585 Japan.
[†††††]. The author is with Panasonic Corporation, Yokohama-shi, 225–8539 Japan.
[††††††]. The author is with Intel Corporation, Turgeneva str., 30, Nizhny Novgorod, Russia.
[†††††††]. The author is with NTT DOCOMO, INC., Kanagawa, 239-8536 Japan.
[††††††††]. The author is with Toyota Motor North America, Inc., Mountain View, CA, 94043 USA.

Millimeter wave communication (mmWave) will be a key component in supporting enhanced use cases of automated driving. The main motivation for mmWave frequencies is the higher bandwidths available at those frequencies [12][13], and thus higher potential data rates. For this reason, it has been adopted for local area networking in IEEE 802.11ad with peak data rate of 6.75 Gbps [14] and also for 5G NR (New Radio) with peak data rate of 20 Gbps [15]. More recently, mmWave has been proposed as a means to enable raw sensor data sharing in V2X [16][17]. Recent research in [18][19] has derived a required data rate for V2X to ensure safe automated driving, which shows that there is a need for of multi-gigabit communication in V2X use cases. As a result, research has targeted specific challenges related to mmWave for V2X such as position-aided beam training via machine learning [20] and sensor-aided beam training [21][22]. IEEE is now specifying next generation V2X standard of IEEE 802.11bd [23] as a successor of IEEE802.11p and IEEE 802.11ad, and also 3GPP is now specifying NR-V2X [24][25][26] as an adoption of NR in V2X.

This article is devoted to understanding the technical background of mmWave V2X for the support of automated driving. We start by providing existing and more advanced use cases of V2X in Sect. 2. Then, the paper provides background on V2X technologies by explaining the state-of-the-art of IEEE 802.11p and LTE-V2X in Sect. 3. We then present an overview of mmWave V2X for cooperative perception in Sect. 4. We provide calculations to argue why cooperative perception is important and to motivate the high data rates required. We then explain how IEEE and 3GPP technologies are being evolved to support mmWave V2X. Finally, several projects for intelligent street are introduced to show the technical feasibility of mmWave V2X.

## 2. V2X for Automated Driving

This section introduces existing and advanced use cases of V2X. Sect. 2.1 provides existing use cases designed to assist human drivers, while advanced use cases for automated driving are described in Sect. 2.2.

2.1 Existing V2X Use Cases

Future vehicles will be equipped with increasing computational capabilities and various vehicular sensors, enabling intelligent vehicles to perceive their surroundings. V2X is envisioned as a complementary vehicular sensor with a unique potential to detect other vehicles or vulnerable road users, even without a line-of-sight (LOS) between the road users. This capability enhances the perception of intelligent vehicles significantly and enables nearby vehicles to receive relevant data for informed decisions of safety-related applications.

The first generation of V2X was introduced with IEEE 802.11p and the exchange of either periodic messages or event-driven messages. SAE Basic Safety Messages (BSMs) and the European Telecommunications Standards Institute (ETSI) Cooperative Awareness Messages (CAMs) are periodic broadcast messages with information about position, speed, heading, etc. [1]. Differently, ETSI Decentralized Environmental Notification Message (DENM) are event-driven broadcast messages [2] for V2V communication. V2I messages broadcasted by RSU are for instance Signal Phase and Timing (SPAT) to provide the traffic light signal phase and timing, and Map Data (MAP), which provide the topology/geometry of a set of lanes at an intersection [4]. All of these message types are also used for V2X based on LTE.

The 3GPP use cases described in TR 22.885 "Study on LTE support for V2X services" [5] are categorized into different types of vehicular communications: Vehicle-to-Vehicle (V2V), Vehicle-to-Infrastructure (V2I), Vehicle-to-Network (V2N) and Vehicle-to-Pedestrian (V2P). The use cases for V2V involve warning applications such as forward collision warning, control loss warning, emergency vehicle warning, queue warning, wrong way driving warning, emergency stop warning and pre-crash sensing warning. Within these scenarios, BSMs, CAMs, or DENMs are transmitted to warn the driver, who then reacts to the dangerous traffic situation accordingly. In addition, these messages can also be used to directly act on the vehicle control system and prevent accidents through a direct influence on the decision-making entity in an automated vehicle. Another V2V use case in [5] describes platoon management, a scenario where a group of vehicles uses V2V to control the speed collectively. This V2V use cases provides convenience and safety benefits to the participating vehicles and is also highly relevant for automated driving applications. The V2I/V2N use cases in [5] describe road safety services via infrastructure, for instance through optimization of traffic flow for vehicles approaching an intersection, which is also highly relevant for automated driving applications. As shown in Fig. 1, this scenario describes an application that is able to fully replace traffic signs by controlling the speed of the approaching vehicles.

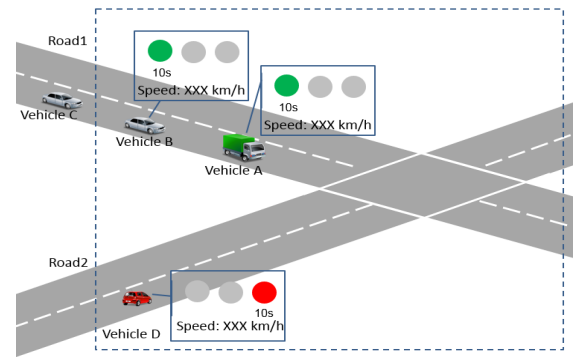

**Fig.1** Traffic Flow Optimization (source:[5])

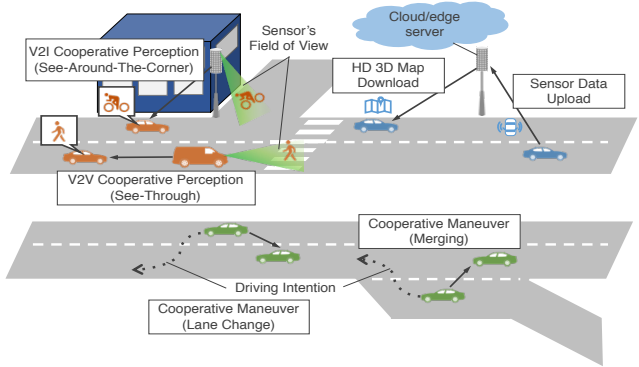

**Fig. 2** Examples of advanced V2X use cases for automated driving.

2.2 Advanced V2X Use Cases

More advanced V2X use cases are described in the 3GPP document TR 22.886 "Study on enhancement of 3GPP Support for 5G V2X Services" [6]: Vehicles Platooning, Extended Sensors, Advanced Driving and Remote Driving. To enable these next generation V2X applications, the transmission of BSMs and CAMs will have to be complemented with transmission of larger messages containing processed sensor data, raw sensor data, vehicles' intention data, coordination, confirmation of future maneuver etc. For these "enhanced V2X" (eV2X) use cases, the requirements on data rate, reliability, latency, communication range and speed are far more demanding.

Automated driving applications with higher Levels of Automation (LoA) further increase these strict requirements. In SAE J3016 [27], a distinction is drawn between LoA 0-2 (Driver Control), where the human driver is primarily responsible, and LoA 3-5 (Vehicle Control), where the automated system is primarily responsible for monitoring the driving environment. For instance, the Extended Sensors use case under the assumption of Driver Control requires an end-to-end latency of 100 ms, whereas the same use case under the assumption of automated driving requires an end-to-end latency of 3 ms [6]. To meet the requirements of V2X use cases for automated driving, one has to understand in detail the participating parties, the specific environment and the particular goal of these eV2X uses cases.

Figure 2 shows some examples of advanced V2X use cases for automated driving, described in more detail below:

*Cooperative maneuver*: This enables maneuver coordination by exchanging some information among automated vehicles in proximity, e.g., for safe, efficient operation of lane change and merging on a highway, at an intersection, or at a roundabout. Recently, SAE J3216 [28] defined four classes of cooperative driving automation, depending on cooperation capabilities and types of exchanged information:

- Class A (status-sharing cooperation): Perception information about the traffic environment and information about the sender (i.e., "Here I am, and here is what I see.") for LoA 1 through 5
- Class B (intent-sharing cooperation): Information about planned future actions of the sender (i.e., "This is what I plan to do.") for LoA 1 through 5
- Class C (agreement-seeking cooperation): A sequence of collaborative messages among specific vehicles intended to influence local planning of specific driving actions (i.e., "Let's do this together.") for LoA 3 through 5
- Class D (prescriptive cooperation): The direction of specific action(s) to specific traffic participants, provided by a prescribing entity (e.g., emergency vehicles) and adhered to by a receiving entity (i.e., "I will do as directed.") for LoA 3 through 5

*Cooperative perception*: This enables sharing the processed (e.g., detected object information, referred to as Collective Perception Messages (CPM) [29]) or raw onboard sensor data, e.g., camera, Light Detection And Ranging (LiDAR), radar, between vehicles in proximity and/or with infrastructures with low latency for safe, efficient, and proactive driving. With this capability, vehicles can see through other non-connected vehicles and pedestrians and see around corners at intersections, which onboard sensors cannot see due to sensor occlusion by surrounding objects (e.g., buildings, vehicles, trucks, trees). For example, in the see-through application, a vehicle can obtain the front camera data of the vehicle ahead to recognize the non-line-of-sight (NLOS) traffic situation ahead using V2V communication. In the see-around-the-corner application, infrastructure equipped with various sensors at intersections can distribute real-time sensor data with nearby vehicles using V2I communication, so that the vehicles can recognize the NLOS situation of the perpendicular streets. Also, cooperative perception is useful for platooning to share onboard sensor data among vehicles in the same platoon group for its safe operation. This use case will be featured in Sect. 4 with more detail to describe the necessity of mmWave for V2V and V2X.

*High definition (HD) 3D map download*: A HD 3D map is a key component of accurate localization in automated driving systems. For safe and efficient driving, automated vehicles need to have up-to-date HD 3D map, but the map can become outdated for various reasons such as road construction. In this use case, automated vehicles download the up-to-date HD 3D map data from the infrastructure on-demand. This makes it easy to keep HD 3D map data updated and allows it to reflect changes on much shorter timescales.

*Sensor data upload*: In this use case, vehicles upload the onboard sensor data to the cloud and/or edge servers via infrastructures. The uploaded sensor data can be utilized for constructing crowdsourced HD 3D map data and also traffic analysis at the cloud/edge servers.

## 3. Existing V2X Standards and Current Status

This section introduces two existing V2X standards, i.e. IEEE 802.11p in Sect. 3.1 and LTE-V2X in Sect. 3.2. Latest enhancements are also provided as additional information.

3.1 IEEE 802.11p

IEEE 802.11p [7] is a standardized communication technology for PHY and MAC layers for Wireless Access in Vehicular Environments (WAVE) [30] to support basic safety and non-safety V2X applications. This technology is the basis of Dedicated Short-Range Communication (DSRC) in the U.S. and ITS-G5 in Europe, which operates in the 5.9 GHz ITS band, and ITS Connect in Japan, which operates in the 760 MHz ITS band [31]. The automotive industry has worked on the development, standardization, and commercialization of IEEE 802.11p and upper layer protocols for more than a decade in several international and regional standardization development organizations such as IEEE, SAE, ETSI, ARIB, etc. In addition, IEEE 802.11p was verified with various testing and large-scale field trials [32] by the automotive industry and thus is a mature technology. Indeed, the commercial deployment had already started in Japan from 2015, in U.S. from 2017, and in Europe from 2019 [33].

IEEE 802.11p can support various basic safety and non-safety V2X applications such as collision avoidance, emergency vehicle notification, vulnerable road user safety, cooperative adaptive cruise control, traffic signal change advisory, red light caution, etc. [31]. The basic process of collision avoidance using IEEE 802.11p is that every vehicle supporting IEEE 802.11p simultaneously 1) broadcasts its state information (e.g., the location, speed, acceleration, and heading) several times per second (typically 10 times per second) over a range of several hundreds of meters and 2) receives these messages from surrounding vehicles to assess collision threats based on both its own state information and also on shared state information from other vehicles. When the onboard system detects a collision threat, the system warns the driver to avoid the potential collision.

Figure 3 illustrates the protocol stack of DSRC used in the U.S., where IEEE 802.11p is employed for the PHY and MAC layers. DSRC is based on IEEE 802.11a with several modifications for more robust operations in vehicular environments. On the PHY layer, just as in IEEE 802.11a, Orthogonal Frequency Division Multiplexing (OFDM) with convolutional coding is employed. Different from the 20 MHz in IEEE 802.11a, IEEE 802.11p has a 10 MHz channel width, and doubled guard intervals, training sequences, and preambles to counter the impact of larger delay spread in outdoor vehicular scenarios, which are severer than one in Wireless Local Area Network (WLAN) environments (e.g., indoor environments). At the MAC layer, IEEE 802.11p inherits Carrier Sensing Multiple Access with Collision Avoidance (CSMA/CA) mechanism. To enable low latency communications and bypass the time-costly procedure of setting up Basic Service Sets (BSS) required in the traditional 802.11 family of WLAN protocols, IEEE 802.11p uses new lightweight rules called "Outside of the Context of BSS (OCB)", which allow vehicles to transmit signals without prior association. Also, to mitigate packet collisions in congested scenarios, decentralized congestion control techniques were defined, which adapt the message interval and transmission power to react to channel congestion.

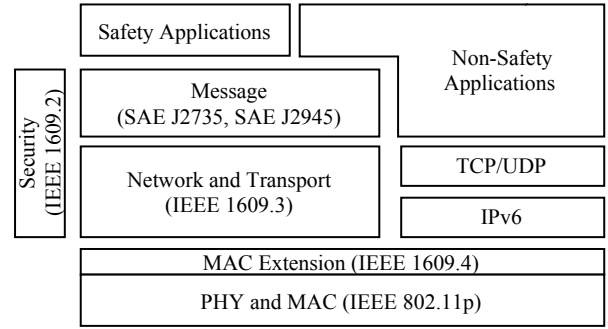

**Fig. 3** Protocol stack of IEEE 802.11p-based DSRC used in the U.S

IEEE 802.11 Working Group established a new task group, TGbd, for next generation V2X (NGV) and the task to develop a new amendment IEEE 802.11bd as the successor of IEEE 802.11p. IEEE 802.11bd targets communications in the 5.9 GHz band and optionally in the mmWave band from 57 GHz to 71 GHz. IEEE 802.11bd will have interoperability, coexistence, backward compatibility, and fairness with IEEE 802.11p devices. The target performance is to achieve [34]:

- At least 2 times higher throughput at MAC layer than the maximum mandatory data rate of IEEE 802.11p defined in the 5.9 GHz band (12 Mbps in a 10 MHz channel) at vehicle speeds up to 250 km/h (closing speeds up to 500 km/h)
- At least 3 dB lower sensitivity level (i.e. larger communication ranges) than that of the lowest data rate defined in IEEE 802.11p in the 5.9 GHz band (3 Mbps in a 10 MHz channel)
- At least one form of positioning in conjunction with V2X communications.
- The specification of IEEE 802.11bd in the mmWave band is still under development in IEEE 802.11 TGbd. In [35], it was proposed to reuse the PHY and lower MAC layers of IEEE 802.11ad with some specification changes for V2X communications. One of expected specification changes from IEEE 802.11ad is to introduce an OCB mode to cope with rapidly changing communication environments in vehicular scenarios. In Sec. 4.3, to examine the applicability of IEEE 802.11ad for V2X communications, we show experimental evaluation results of IEEE 802.11ad in outdoor V2I scenarios.

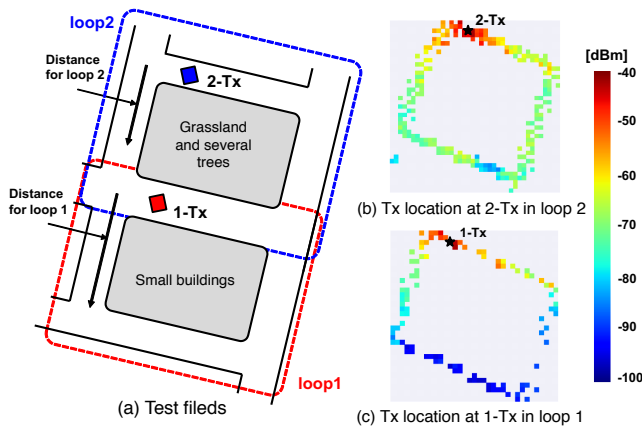

Fig. 4  Example of measurement-based spectrum database observed by field test in 5.9GHz band

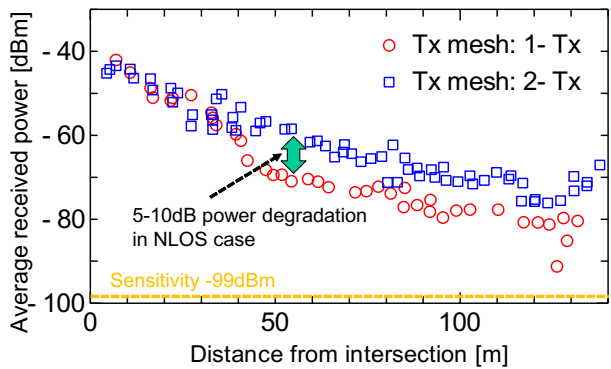

Fig. 5  Predicted received power with different Tx positions.

Table I  LTE V2X requirements

| | |
|---|---|
| Latency | Maximum of 100 ms for V2V, V2I, V2P (maximum of 20 ms for particular V2V usage, i.e. pre-crash sensing). Maximum of 1000 ms for V2N (Vehicle-to-Network) |
| Message payload size | 50 – 300 bytes for periodic broadcast messages between two UEs. Up to 1200 bytes for event-triggered messages between two UEs. (Note that above values do not include security-related message component) |
| Data Rate | Maximum frequency of 10 messages per second for one transmitting UE (maximally 24 kbps) |
| Velocity | UE's maximum absolute velocity of 250 km/h |

**Reliability Improvement for Future DSRC**

Big data technologies can be used for improving the reliability of distributed systems. For examples, a smart spectrum concept proposed in [36] can use wireless big data for a flexible spectrum management of future mobile communication systems. The spectrum database is established by using radio environment measurement data from mobile terminals located all over the world, so-called crowd sensing. V2V systems are considered to be a suitable application since an antenna can be mounted on the rooftop of a vehicle and stable spectrum measurement results can be obtained. The measured data is gathered at the server and processed statistically to predict the propagation between the transmitter location and the receiver location. The database can be installed at a cloud network or edge of the network [37]. By using this database for predictions of the radio propagation, quality of V2V communication with respect to radio propagation can be understood before entering the specific scenarios such as intersection.

A field test for predicting the received power of V2V communication by using measurement-based spectrum database in the 5.9 GHz band was introduced in [38]. In this field test, three vehicles were equipped with an on-board unit, IEEE 802.11p-based V2X radio. Received Signal Strength Indicator (RSSI) and location of both transmitter and receiver information are first stored in a local storage and then uploaded to the spectrum database. At the database, the measurement data is averaged within a certain square mesh for each transmitter location and receiver location. We select a test field in California PATH, Richmond CA, U.S.A. Fig.4 (a) shows the test fields of the measurement campaign. There are many small buildings in loop 1 and only a few in loop 2. Therefore, we can compare the radio propagation of Line of Sight (LOS) in loop 2 and Non-Line of Sight (NLOS) in loop 1. Figs. 4 (b) and (c) show examples of radio environment maps with transmitter positions in loop 1 and loop 2 within 5 m mesh square, respectively. It can be observed that the influence of the building is reflected in the data and a prediction of the received power can be performed if the transmitter and the receiver locations are known. Fig.5 shows the average received power with different Tx positions in 1-Tx and 2-Tx. The horizontal axis shows the distance from intersections in loop 1 and loop 2 shown in Fig. 4 (a). We can understand that the average received signal power transmitted from mesh 1-Tx and mesh 2-Tx are different tendency due to the effect of buildings. From this figure, we can confirm that the measurement-based spectrum database by using crowd sensing visualizes the influence of obstacles from the transmitter and the receiver.

In this example, we show the case of the received signal power prediction by using crowd sensing, but we can also use crowd sensing to predict other characteristics such as packet loss rate and throughput. The measurement-based spectrum database is effective in higher frequency because the influence of deep shadowing can be predicted. While the crowd sensing is not a function defined in IEEE802.11p/bd, it can be used for transmission power control, adaptive modulation and coding to improve reliability of V2X in the future. The crowd sensing is a concept independent from wireless standards and defined in a higher layer. The big data effectively used in V2X are not limited to spectrum information [37][39]. If we can gather the information of suitable beam for communication, the reliable beamforming can be designed. We can also obtain higher layer information such as latency of network, vehicle traffic, road situation and weather information. Moreover, the future connected vehicles with big data have an important role of information and computation infrastructure such as Intelligent Internet of Vehicle (IoV) concept summarized in [37].

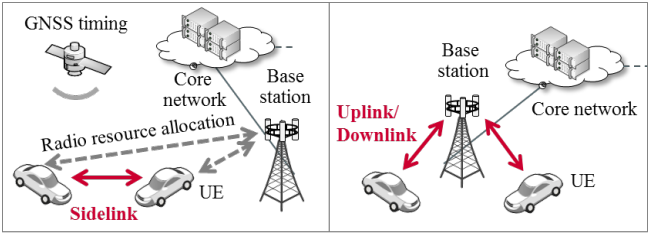

**Fig.6** Sidelink, uplink and downlink communication in LTE V2X

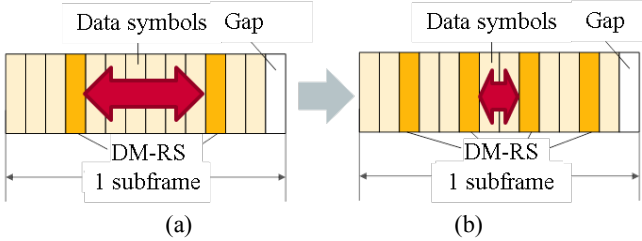

**Fig.7** Frame structure of sidelink based LTE V2X

3.2 LTE V2X

LTE is currently deployed on a wide scale and provides services over two hundred of countries in the world [40]. 3GPP has initiated LTE-based V2X standardizations in Release 14 [10][11]. The aim of these 3GPP activities is to enhance LTE systems and enable vehicles to communicate with other vehicles, pedestrians, and infrastructures in order to exchange messages for road safety, controlling traffic flow, and providing various traffic notifications. LTE has the potential to support various V2X services due to its widely deployed network and User Equipment (UE), secured network, high spectrum efficiency, wide coverage, high mobility, high reliability, low latency, long battery life, and so on. In 3GPP, several use cases and associated requirements are identified [41] as shown in Table I by referring the existing V2X message sets which are specified outside 3GPP [3].

Since the first standardization of LTE (Release 8) in 2008, LTE was continuously evolved over several releases by 3GPP. This evolution did not only include DL and UL enhancements between base stations and UE, but also covers multiple types of communications. For example, Single-Cell Point-To-Multipoint (SC-PTM) is a broadcast/multicast service from one single cell to multiple UEs [42] and Multimedia Broadcast Multicast Services (MBMS) is a broadcast/multicast service from multiple cells to multiple UEs [43].

In addition, 3GPP specified LTE sidelink communication. Sidelink communication corresponds to direct communications between UEs as shown in Fig.6. Compared with IEEE 802.11p, LTE sidelink communications has some technical differences such as Hybrid Automatic Repeat reQuest (HARQ), scheduling, Turbo coding for the data channel. For the LTE V2X sidelink specifications [10], some enhancements are introduced as follows.

**Out of network coverage operation**

Similar to the Release 12 LTE sidelink specifications, radio parameters of UEs are pre-configured for out of network coverage operation where no base stations to control the sidelink based V2X communications are deployed (transmission mode 4). The vehicles autonomously select resources for their sidelink transmissions based on the distributed scheduling scheme. A remote provision of pre-configured parameters via the LTE network is being considered.

**Frame structure**

In LTE, a subframe of 1 ms length consists of 14 symbols. In the Release 12 sidelink specifications, two DeModulation Reference Signals (DM-RS) are time-multiplexed in a subframe as shown in Fig.7(a), where the last symbol is used for Tx/Rx switching and timing adjustment. To handle the high Doppler frequency associated with relative speeds of up to 500 km/h at high carrier frequency (e.g. 5.9 GHz), additional DM-RS symbols have been added as shown in Fig.7(b) to achieve better tracking of the channel at the high speed.

**Resource allocation**

The UL frame structure is utilized to distribute shared radio resources. Base station-controlled resource allocation and UE autonomous resource allocation are both supported. Semi-persistent scheduling with background sensing is introduced for the UE autonomous resource allocation. A transmitter will select less interfered radio resource based on the detected control information and measurement. Control information also includes reservation indication for the semi-persistent scheduling transmissions. A sensing window prior to resource selection is used to identify occupied resource in the resource candidate for transmission. Transmission resource will be selected within a certain selection window size which is determined based on the latency requirement.

**Synchronization**

Generally, the location information can be obtained from a local Global Navigation Satellite System (GNSS) receiver, which also allows for a GNSS-based synchronization among UEs. Synchronization signal transmitted from base station and/or neighbor UEs can also be used for synchronization.

**Enhancement of LTE V2X in Release 15**

To further enhance the Release 14 LTE V2X technologies and to support more advanced vehicular applications, several new V2X features are introduced in Release 15. For example, the support of carrier aggregation expands the UE bandwidth by employing radio resources across multiple component carriers, which is particularly advantageous for the sidelink and its new resource allocation procedure. Other important new features are a higher modulation-level, a reduced latency, and the transmission diversity.

## 4. Towards mmWave V2X

This section gives an overview of mmWave V2X for supporting enhanced V2X use cases for automated driving. Sect. 4.1 provides calculations to argue why cooperative perception is important and to motivate the high data rates required. Technical feasibility using mmWave V2X are evaluated in Sect. 4.2 under basic V2X scenarios. How IEEE and 3GPP technologies are being evolved to support mmWave V2X is explained in Sect. 4.3 and 4.4 respectively. Finally, Sect. 4.5 introduces several projects for intelligent street to show the feasibility of mmWave V2X in real environments.

4.1 mmWave V2X for Cooperative Perception

MmWave V2X is suitable to complement IEEE 802.11p and LTE V2X because of its high-data-rate capability. Due to the limited bandwidth, IEEE 802.11p and LTE V2X can support data rate of up to 27 Mbps and 28.8 Mbps respectively. On the other hand, mmWave V2X can support a data rate of up to 1 Gbps or more, thanks to the wide bandwidth available in mmWave. Therefore, mmWave V2X is beneficial in supporting advanced V2X use cases (e.g. camera/LiDAR sensor data sharing) that require higher data rates.

Automated driving systems require high resolution and real-time maps, so-called dynamic HD maps, to maneuver vehicles safely [44]. A LiDAR (Light Detection and Ranging) sensor is typically used to generate HD maps and to monitor the vehicle surroundings, which can then be displayed as a high-resolution and real-time point cloud. However, in complex urban city environments, the visible area of a HD map from an ego perspective can easily be blocked by other vehicles, surrounding buildings, and parked vehicles along the street. This blocking problem and the hidden objects are a challenge for safe automated driving and degrades efficiency of automated driving due to a limitation of the maximum allowable velocity of vehicles.

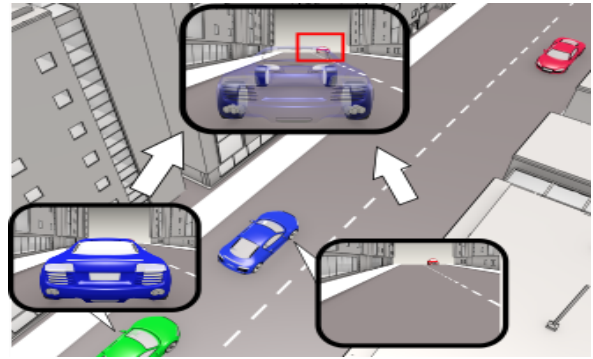

**Fig. 8** Cooperative perception by using extended sensor sharing between ego vehicle (green) and blocking vehicle (blue) to detect oncoming vehicle (red).

Cooperative perception is realized by exchanging sensor data between vehicles or/and RSUs, which is necessary to widen or enhance the visibility area of HD maps [45][46]. Fig. 8 shows an example of cooperative perception in an overtaking scenario where an ego vehicle (green) is trying to overtake a blocking vehicle (blue) while an oncoming vehicle (red) is coming. As indicated in the Fig. 8, the visible area of the ego vehicle is completely blocked by the blue vehicle, while the visible area of the blocking vehicle is able to detect the oncoming vehicle. Therefore, exchanging sensor data from the blocking vehicle to the ego vehicle is effective in detecting hidden objects and make it possible to realize cooperative perception. It is important to note that low latency is a crucial system parameter for sensor data exchange in order to realize stable control. It is well known that latency and data rate are a tradeoff for the case of video or LiDAR data transmission. The latest video data compression techniques reduce the data rate to one-tenth of the raw data rate. However, such video compression leads to latencies much larger than the 3 ms required for automated driving. Therefore, raw (or nearly raw) sensor data will likely be needed to ensure latency requirements, which also adds value for liability investigations in case of an accident. To realize cooperative perception via V2V communication, we need to clarify the requirements on data rate to ensure safe automated driving. The main objective in this subsection is the derivation of these requirements, which will show that mmWave in V2V communications is feasible and reasonable.

The overtaking scenario in Fig. 8 is targeted problem in this subsection, which is described in more detail in Fig. 9. The ego vehicle is located behind the blocking vehicle with an inter-vehicle distance of $d_{be}$, while the oncoming vehicle has a distance $d_{oe}$ to the ego vehicle. It is assumed that all vehicles are driving with the same velocity $v$. It is also assumed that the ego and blocking vehicles are equipped with LiDAR sensors, while the oncoming vehicle is not equipped with any sensors and disconnected from vehicular networks. Laser beams of the LiDAR from the blocking vehicle are approaching the oncoming vehicle (cyan), while beams from the ego vehicle are blocked by the blocking vehicle (yellow).

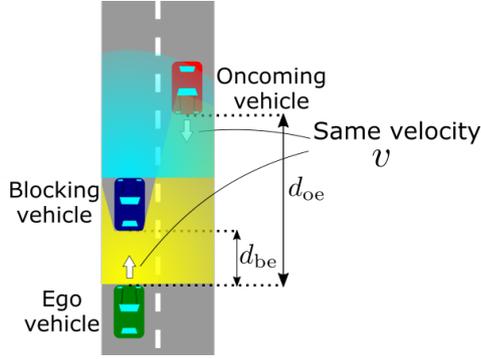

**Fig. 9** Overtaking scenario of ego vehicle against blocking vehicle while oncoming vehicle is coming.

During the moment of the overtaking process where the ego vehicle detects the oncoming vehicle, the distance $d_{oe}$ should be larger than twice the braking distance $d_{brake}$ to prevent a crash of the ego vehicle with the oncoming vehicle. In case of comfortable braking and for typical road conditions, the braking distance can be described as in Eq. (1) [47].

$$d_{brake} = 0.039 \frac{v^2}{3.4} \quad (1)$$

In general, the data rate of a LiDAR sensor (transmitting from the blocking vehicle to the ego vehicle for cooperative perception) can be described by Eq.(2), where $B_{laser}$ is the number of bits per laser, $T$ is the scan period of the LiDAR, and $N$ is the number of LiDAR lasers needed to detect the oncoming vehicle.

$$R_{req}(v, d_{be}) = \frac{B_{laser} N_{req}(v, d_{be})}{T} \quad (2)$$

The minimum required number of lasers $N_{req}$ can be defined as in Eq. (3), where $\tilde{N}(d_{oe}, d_{be})$ is the number of reflected lasers from an unit surface on the target object (oncoming vehicle) with and without cooperative perception and $\tilde{N}_{th}$ is detection threshold, which is determined by the minimum required density of reflected lasers needed to detect objects. It is to be noted that the required data rate $R_{req}$ becomes a function of the velocity $v$ and the inter-vehicle distance $d_{be}$.

$$N_{req}(v, d_{be}) = \arg\min N$$
$$\text{s.t. } \tilde{N}(d_{oe}, d_{be}) \geq \tilde{N}_{th} \text{ when } d_{oe} = 2 \times d_{brake}(v) \quad (3)$$

Figure 10 shows the required data rate on V2V communication in terms of the vehicle velocity $v$ at the inter-vehicle distance $d_{be} = 10$ m, derived by using the theory described in this subsection. The blue line in Fig. 10 shows that without cooperative perception but only ego perception (blue line), there is a limitation on the velocity regardless of the sensor data rate. On the other hand, cooperative perception (red line) improves efficiency of driving, since a higher data rate

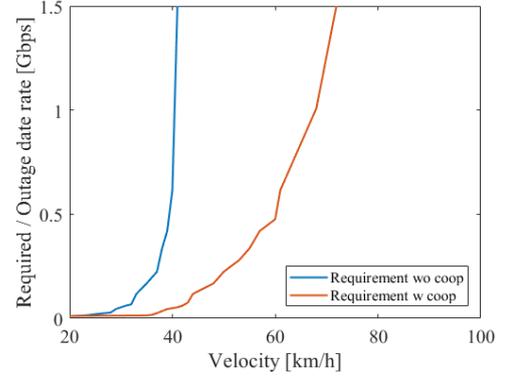

**Fig. 10** Required data rate to ensure safe driving at inter-vehicular distance $d_{be} = 10$ m with different vehicle velocities (blue: without cooperative perception using only ego sensor, red: with cooperative perception using extended sensor)

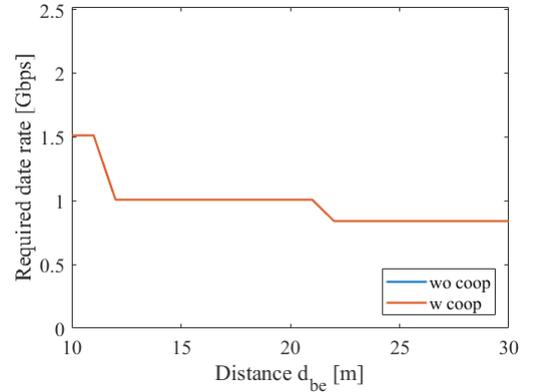

**Fig. 11** Required data rate to ensure safe driving at a vehicle velocity $v = 70$ km/h with different inter-vehicular distance (blue: without cooperative perception, which is impossible to ensure safety driving, red: with cooperative perception using extended sensor)

increases the allowable maximum velocity of safe automated driving. It is noted that the blue line is showing the required sensing rate of the LiDAR equipped on the ego vehicle, while the red line is showing both the required sensing rate of the LiDAR equipped on the blocking vehicle and the required communication rate between the blocking and ego vehicles to transfer the sensing data. For the cooperative perception case, it can be found that a data rate of at least 1 Gbps is needed to ensure safe automated driving with a vehicle velocity of 70 km/h. This fact is the motivation to introduce mmWave for V2X.

Figure 11 provides another analysis on the required data rate with respect to inter-vehicle distance and a fixed vehicle velocity of $v = 70$ km/h. Although the required data rate becomes smaller with increasing the inter-vehicle distance, it is found that a data rate of around 1 Gbps is required for a distance of up to a few tens of meters, as can be seen in

Fig.11. These requirements are the reason for a study on enhanced V2V in [17] and also in 3GPP [48], which will be explained in more detail in the later subsections.

4.2 Technical Feasibility using mmWave V2X

This subsection provides data rates realized with mmWave V2X, which justifies that mmWave is the only feasible technology for safe automated driving. Based on the calculated data rate, the maximum allowable velocity of vehicles to ensure safe automated driving is evaluated. It also describes specific challenges related to mmWave for V2X including coverage enhancement and beam alignment.

Figure 12 illustrates an example of a V2V communication link. For a simple analysis, the two-path channel model is introduced, where the direct wave has a path length $r_d$ while the ground reflected wave has a length $r_r$. To overcome fading effect caused by two-path propagation and fluctuating vehicle positions, antenna selection diversity is introduced at the receiver side with different antenna heights $h_{r1}$ and $h_{r2}$.

As described in Table II, three different frequency bands of 5, 30, and 60 GHz are introduced for V2V with different antenna gain and bandwidth. The antenna spacing for selection diversity changes also depending on the wavelength of each frequency. A higher carrier frequency has a higher antenna gain and a higher bandwidth, which will contribute to the enhancement of safe automated driving, even if the higher frequency is suffering from a larger propagation loss and a larger noise power.

The 0.01% outage data rate (99.99% reliability rate) are calculated based on the channel capacity and plotted in Fig.13 for different frequency bands, both with and without antenna selection diversity. First of all, it is found that the effect of selection diversity is significant and can reduce the performance degradation due to deep fading from two propagation paths and fluctuation of vehicle positions. Secondly, it becomes clear that higher frequency can improve the outage data rate, due to a sufficiently high antenna gain to compensate the propagation loss and the large noise power. Finally, the calculated outage data rates are compared with the required data rate in Fig.14 at an inter-vehicle distance of 10 m. As a result, it can be found that 30 GHz and 60 GHz V2V allow maximum velocities of about 117 km/h and 158 km/h respectively, while 5 GHz V2V allows only 49 km/h. This confirms that mmWave V2V is needed to realize efficient and safe automated driving in the future.

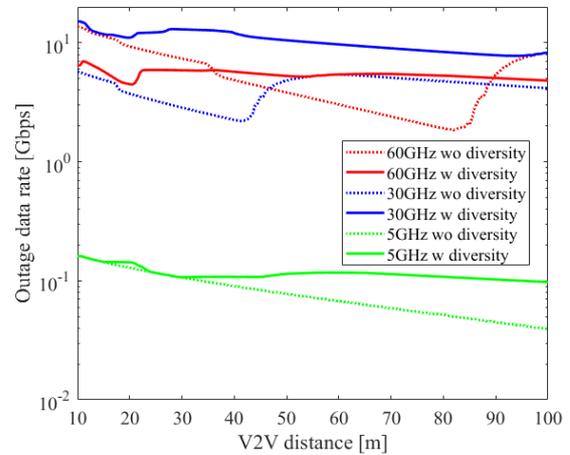

**Fig.13**  The 0.01% outage data rate of V2V with respect to inter-vehicle distance, with and without antenna selection diversity (green: frequency band of 5 GHz, blue: frequency band of 30 GHz, red: frequency band of 60 GHz)

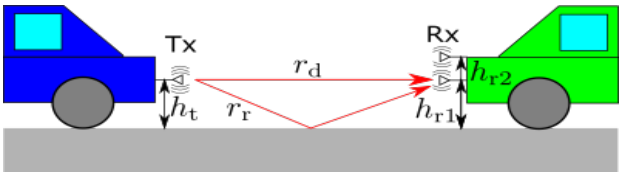

**Fig.12**  Millimeter-wave V2V with antenna selection diversity over two-path propagation channel.

**Table II**  Simulation parameters in milimeter-wave V2V

|  | 5 GHz | 30 GHz | 60 GHz |
|---|---|---|---|
| Transmit power | 10 mW | | |
| Antenna gain | 4.3 dB | 20 dB | 26 dB |
| Polarization | Vertical | | |
| Antenna aperture | 2.5 cm x 2.5 cm | | |
| Antenna diversity | Antenna selection diversity | | |
| Antenna spacing | 120 cm | 20 cm | 10 cm |
| Antenna fluctuation | Gaussian distribution with $\sigma = 3.2$ cm | | |
| Bandwidth | 10 MHz | 500 MHz | 1 GHz |
| # of subcarriers | 512 | | |

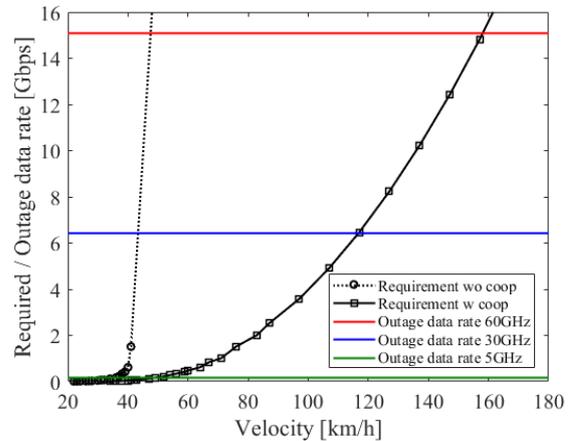

**Fig.14**  Comparison of required data rate and realized outage data rate of V2V (green: frequency band of 5 GHz, blue: frequency band of 30 GHz, red: frequency band of 60 GHz) in terms of velocity of vehicle. If the realized data rate is higher than the required data rate, the V2V communication ensures safe automated driving at that velocity of vehicle.

The benefits of mmWave can also be realized in V2I, where vehicles use infrastructure to relay their messages. The propagation considerations in V2I are different from V2V, due to a different placement of the antennas. In the V2I case, it is more likely that antennas are placed on top of vehicles resulting, compared to the V2V setting, in a more favorable propagation link between the RSU (base station) and the vehicle. In this setting though, buildings and large vehicles become a major impairment. The performance of V2I communication systems in urban areas has been studied in various settings (see for example the review in [48] and the references therein). An analysis of such systems was performed in [49] using a stochastic geometry model inspired by recent measurements [50]. It was found that interference from cross or parallel streets can be neglected due to high losses in these indirect paths. The significance of interference was also found to depend on the street and RSU densities.

Beam alignment is a major challenge in V2I systems. Beam alignment refers to the process of pointing the transmit and receive beams with the goal of achieving the most beneficial array gain. IEEE 802.11ad and 5G use hierarchical search procedures which involve trials of several beam pair candidates and ultimately a choice of the best combination. An alternative for vehicular settings is the exploitation of side information, such as location information, to support pointing of beams. Machine learning can be used to improve robustness of beam alignment for the case when links are blocked by large vehicles. In [51], such a system was developed and called inverse fingerprinting. The key idea is the identification of promising beam pairs as a function of position. A machine learning engine recommends pairs based on location information, which is for example obtained through DSRC. This allows the system to try only a few beam pairs and increases the probability of finding a good pair, instead of trying all possible pairs. An illustration

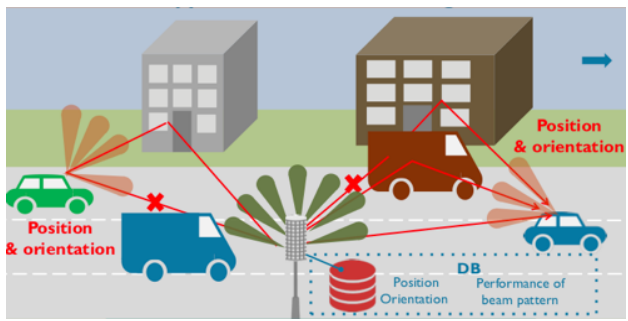

**Fig. 15**  A data driven approach for beam training. Promising beam pairs are identified as a function of location (illustration courtesy of Nuria Gonzalez Prelcic).

of such a system is provided in Fig.15. There may be even more synergies to be exploited. For example, LIDAR data

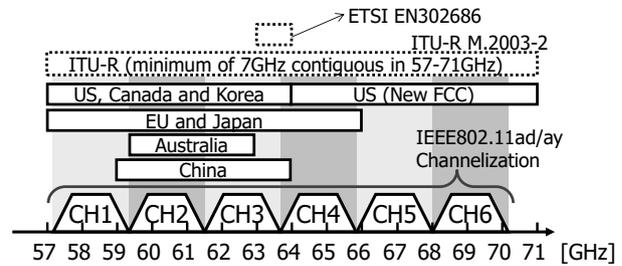

**Fig. 16**  Global frequency allocation of 60 GHz band radio applications

can be used to reduce overheads in mmWave beam training [20].

4.3 IEEE Technology towards mmWave V2X

The frequency band of 60 GHz is a worldwide unlicensed band, and the IEEE 802.11ad standard is recognized as an advanced wireless LAN standard [52] at 60 GHz. IEEE802.11ad provides a data rate of up to 6.75 Gbps with a single channel of 2.16 GHz bandwidth. Commercial products in conformance with the IEEE802.11ad standard have been introduced in various indoor applications such as PCs, tablets, smartphones. ITU-R/WP-5A Recommendation M.2003-2 [53] refers to the IEEE 802.11ad standard as "Multiple Gigabit Wireless Systems in frequencies around 60 GHz". At least 7 GHz spectrum in the 57-71 GHz is allocated to satisfy the requirements of the applications envisioned as shown in Fig.16. For example, the U.S. has regulated 6 channels with each a bandwidth of 2.16 GHz in the 60 GHz band as shown in Fig. 16.

ETSI has specified 63-64 GHz for V2X applications (ETSI EN 302 686 v 1.1.1 [54]). As shown in Fig. 16, 63-64 GHz overlaps with channels 3 and 4 of IEEE 802.11ad, which led to the concern that both systems may interfere with each other. ETSI examined the coexistence of EN302686 and the IEEE802.11ad standard, and the examination result was issued as ETSI Technical Report, TR 103 583v1.1.1 in August 2019 [55]. In this report, 63 to 64 GHz, the frequency for ITS in the 60 GHz band, is recommended to harmonize with the channel arrangement of IEEE802.11ad. In response to the results of this examination and recommendations, there were considerations that it is desirable to move to channel 3 or channel 4 of IEEE 802.11ad even in European Conference of Postal and Telecommunications Administrations (CEPT). It is expected that ETSI will specify channel 3 or 4 of IEEE 802.11ad as V2X applications, so advanced V2X in the 60 GHz band are expected to be utilized in major European countries. Based on such a background, this section discusses the applicability of V2X based on the IEEE 802.11ad protocol, which is the wireless LAN standard in the 60 GHz band.

The IEEE 802.11ad standard was made on the premise of indoor use, but in order to evaluate the applicability to V2X,

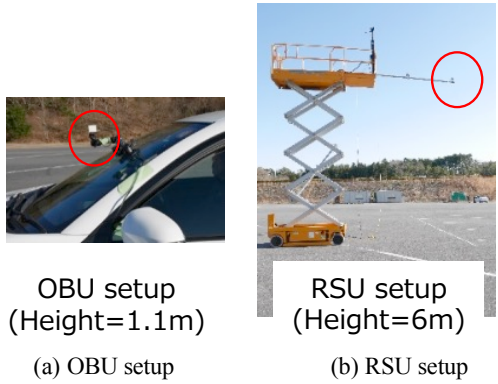
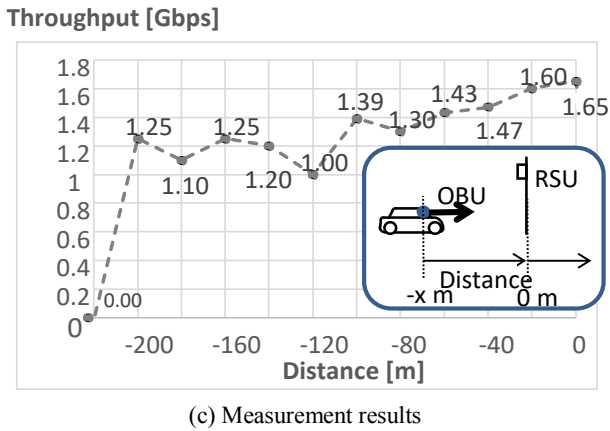

**Fig.17** Measured throughput vs. distance between the RSU and the OBU in stationary (low velocity) state. (a) OBU setup, (b) RSU setup, (c) Measurement results

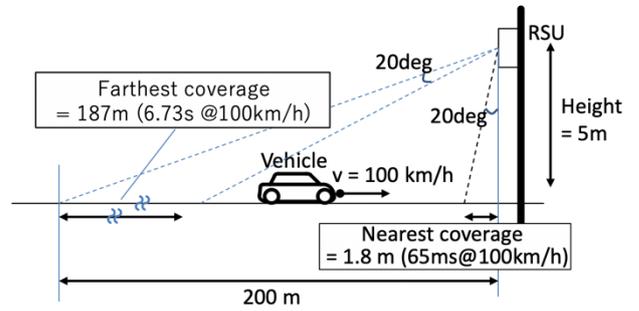

**Fig.18** Conditions for examining the frequency of BFT. RSU height = 6 m, OBU height = 1 m, maximum transmission distance = 200 m

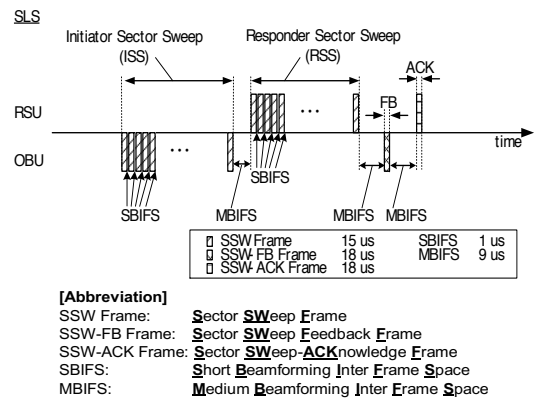

**Fig.19** Beamforming training protocol by Sector Level Sweep procedure

low velocity characteristics in the outdoors have to be evaluated. An RSU was mounted at a height of 6 m and an OBU at a height of 1.1 m. The throughput was measured with respect to the distance between RSU and OBU. Fig.17 shows the measurement results. A throughput of more than 1 Gbps was obtained at all points between 0 m and 200 m [56][57]. In addition to an evaluation with low velocity, the throughput with high mobility was also evaluated. The test showed that 1.09 Gbps can be obtained with 80 km/h velocity at a distance of 100 m, whereas 1.23 Gbps was measured with 90 km/h at 60 m. As described above, although IEEE 802.11ad is a standard for the indoor use case, it becomes clear that IEEE802.11ad is also suitable for outdoor use cases like V2X.

It is further evaluated whether the directional antenna used in IEEE 802.11ad supports mobility. For this evaluation, it is assumed that the height of the RSU is 6 m, the OBU height is 1 m, and the half-power beam width of the antenna directivity is 20 degrees. As shown in Fig.18, for the longest distance at around 200 m, the coverage area of the directive antenna is 187 m, and a connection time of 6.7 seconds can be kept at 100 km/h, which is sufficient time for Beam Form Training (BFT). On the other hand, at the shortest distance to the RSU, the coverage area of the directive antenna is 1.8 m and the connection time is 65 ms. In other words, in order to track the antenna beam, it is necessary to perform BFT at least once every 65 ms. Fig.19 shows the procedure of BFT defined in IEEE 802.11ad. For example, in case of a system with 32 Sector Sweep Frame (SSW-Frame), BFT requires 1.1 ms, which is a sufficiently small value compared to the Beacon Interval (BI) value, generally set to 100 ms. Since IEEE 802.11ad can set BFT twice in a Beacon Interval (BI) of 100 ms, BFT can be executed once every 50 ms. Hence, although transmission efficiency decreases, beam tracking is still possible even at high-speed of 100 km/h.

In this section we evaluated the applicability of IEEE 802.11ad to V2X, mainly focusing on the V2I scenario. Although this standard is defined for indoor usage, it is also capable of handling outdoor usage with a few hundred meters distance and high mobility of around 100 km/h. In future, more technical verification is needed in complicated V2X scenarios to increase the feasibility of V2X using the 60 GHz band.

4.4 3GPP Technology towards mmWave V2X

To meet the requirements associated with the emerging more advanced vehicular applications, particularly for those requiring very high throughputs at a short range, 3GPP has also been exploring the higher mmWave frequency bands for ITS and V2X scenarios, and ultimately evolving LTE-V2X technologies towards NR-V2X.

**Design of New Radio (5G Cellular RAT)**

One of the key attributes and technical advantages of the 5G cellular networks is its natural/inherent capability to efficiently utilize various spectrum. Emerging from the carrier aggregation framework introduced in LTE, the NR system provides flexible options for radio-resource and spectrum management aiming to support different types of devices and applications and enabling communication in various frequency bands. Future NR systems aim to support communication in frequency ranges of up to 100 GHz [58].

The initial version of the NR system design in Release 15 defines only cellular radio-interface (i.e. Uu radio) and supports two Frequency Ranges (FR): FR1 [59] and FR2 [60]. FR1 covers frequencies from 450 MHz to 6 GHz, while FR2 spans the lower part of mmWave bands from 24 GHz to 52.6 GHz (see Table III). The support of further frequency bands is expected to evolve in future releases, and the sidelink air-interface design for V2X use cases will be defined in the 3GPP Release 16 as described in the last part of this subsection. Considering current ITS spectrum regulations in Europe, it can be foreseen that future NR based V2X systems will utilize mmWave band (e.g. 63 GHz) on top of the dedicated ITS spectrum at 5.9 GHz, as well as sidelink communication capability enabled in these frequency bands as well.

MmWave communication includes a powerful framework for active antenna arrays utilizing a combination of analogue and digital beamforming. The support of multi-beam operation is integrated in NR radio-layer procedures starting with initial access and mobility support for DL and UL data communications [61]. Initial access procedure defines periodical broadcasting of beamformed Synchronization Signal Blocks (SSB). When a UE detects SSBs, it determines the system frame timing, acquire the best beam for DL reception and also adjust RX beams. When the UE accesses a NR system by transmitting the Random-Access CHannel (RACH) signal, the association between one or multiple SSBs and a subset of RACH resources and preamble indices helps the network to derive DL beams of SSBs received by UE. The NR SSBs are also used for layer 3 mobility. In particular, the cell level quality is assessed based on SSB Reference Signal Received Power (RSRP) and Reference Signal Received Quality (RSRQ) measurements for multiple spatial beams.

On top of NR SSBs, the Channel State Information Reference Signals (CSI-RS) can be configured to the UE. The CSI-RS signals are user-centric and can provide improved beam-management and mobility performance for connected UEs. Beam management includes multiple procedures such as beam selection, beam measurement and reporting, beam switching, beam indication and recovery [62], which are all designed to ensure optimal TX-RX beam pairs used for DL and UL communication.

Finally, the NR system defines reference signals (DM-RS) used for demodulation of physical control and shared channels. The NR DM-RS density and pattern are reconfigurable to efficiently support scenarios with different mobility assumptions. In addition, the Phase Tracking Reference Signals (PTRS) can be configured to efficiently cope with the phase noise and residual frequency offset/Doppler shift [63].

**Evolution of C-V2X System Requirements**

As discussed in Sect. 3.2, the first generation of V2X system in 3GPP was developed based on LTE technology, where two types of radio-interfaces were supported: cellular (DL/UL) and sidelink. The work on NR-V2X system design has started, motivated by evolved V2X applications and requirements, and identified by the 3GPP system architecture group. The evolved V2X use cases were categorized into four major groups: advanced driving, extended sensors, platooning and remote driving [48]. The radio-layer requirements for eV2X use cases are much more stringent compared to the initial objectives of the LTE-V2X system and will be addressed in a more complete manner in future releases of NR-V2X technology. It is important to note that work on NR-V2X use cases and requirements is still ongoing through close cooperation of the automotive sector (e.g. 5GAA) and other organizations (e.g. SAE, ETSI ITS) that develop standards and promote emerging V2X applications and services.

**Towards mmWave V2X System Design**

The first release of the NR system design enabling cellular communication has been finalized in Release 15 and can provide selected vehicular communication services over the Uu radio interface by enabling communication in both frequency ranges including mmWave (FR2). The NR Uu radio-interface design still has to be further enhanced in order to support multicast/broadcast in DL transmissions, which is a beneficial component for many V2X services.

In Release 16, from physical layer perspective, the NR-V2X sidelink design supports operations in both FR1 and FR2. However, the primary focus for the first generation of the NR-V2X design is on FR1 to support low-latency reliable sidelink communication in low-band ITS spectrum at 5.9 GHz. Different from LTE V2X, the physical layer of NR-V2X technology in FR1 will support new features, such as HARQ for unicast and groupcast communication, multi-layer Multi-Input Multi-Output (MIMO) transmission, enhancements of sensing and resource selection procedures for aperiodic traffic, support of preemption functionality and new physical structure with configurable numerology, and a set of patterns for demodulation of reference signals.

Table III  Definition of frequency ranges (NR R15)

| Frequency range | Corresponding frequency range |
|---|---|
| FR1 | 450 MHz – 6000 MHz |
| FR2 | 24250 MHz – 52600 MHz |

Table IV  FR1 and FR2 Bandwidth Configurations

| SCS kHz | FR1 System Bandwidth, MHz | | | | | | | | | | |
|---|---|---|---|---|---|---|---|---|---|---|---|
| 15/30/60 | 5 | 10 | 15 | 20 | 25 | 30 | 40 | 50 | 60 | 80 | 100 |
| SCS kHz | FR2 System Bandwidth, MHz | | | | | | | | | | |
| 60/120 | 50 | | 100 | | | 200 | | | 400 | | |

The NR-V2X design beyond Release 16 is expected to evolve further. The enhancements of Uu interface and the NR sidelink air-interface are likely to be introduced for communication in mmWave bands and designed respectively to support advanced V2X applications. At the same time, it is also understood that the standalone sidelink V2X communication in mmWave band is technically challenging, including the frequency offset of up to 0.3 ppm (with maximum vehicle relative speed of 500 km/h) and the beam alignment of mmWave links in dynamic V2X scenarios.

4.5 Intelligent Street with mmWave V2X

The integration of mmWave V2X infrastructure into the urban landscape requires a paradigm change and a new approach to wireless network infrastructure. The reason for this paradigm change lays in the physical properties of mmWave, i.e. the increased attenuation at higher frequencies and the requirement to establish a line-of-sight between communicating parties. Due to propagation similarities of mmWave and visible light, it appears logical to mount mmWave infrastructure to streetlamp poles. The availability of electricity and the close position above the mobile users are beneficial and in addition enable an immense opportunity for entirely new services. In fact, several consortia regard intelligent streetlamps, as shown in Fig.20, equipped with smart lighting, video cameras, air pollution sensor and other sensors, as the key enabler and digital backbone for Smart Cities. Examples of relevant projects are the Lightpole Site by Ericsson/Philipps [64], the Smart Pole Pilot in San Jose [65], the 5G BERLIN initiative [66] and the LuxTurrim5G project in Finland [67], the ITS Connect in Japan [68], and the Super Smart Society (SSS) project in Tokyo[69]. While all of the above projects connect lighting poles to a wireless network and regard street lamps as a platform for new digital services, particularly 5G BERLIN, LuxTurrim5G and SSS involve mmWave technology and are therefore able to provide the large bandwidths required to support new eV2X use cases such as HD 3D Map download, sensor data upload for cooperative perception, which ultimately allow for cooperative maneuvers of automated vehicles (see Section 2). Other similar and promising use cases are "3D video composition" [63] or "Massive Automotive Sensing" in [16].

Figure 21 shows the potential of streetlamps equipped

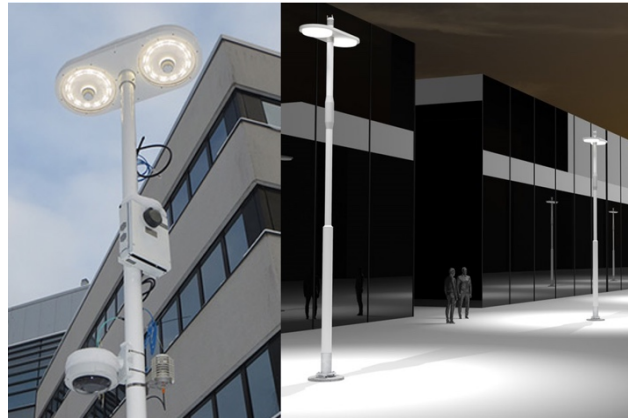

**Fig.20** Smart Light Poles from LuxTurrim5G project

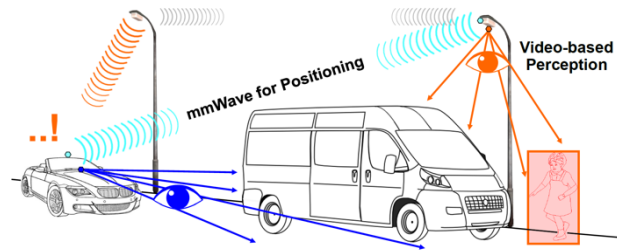

**Fig.21** mmWave V2X for highly accurate positioning and enhanced perception for Automated Driving with mmWave V2X for highly accurate positioning and enhanced perception. The integration of high-resolution video cameras in intelligent streetlamps allow for a bird's eye view and a fundamental advantage compared to the low height of sensors mounted on vehicles. Due to the perspective from above, video sensors mounted on lamp poles allow for a more accurate localization of objects. More important, the bird's eye view allows for a complete perception of traffic situations, even under highly complex circumstances in urban traffic scenarios. Different from the low height of sensors mounted on the vehicle, a sensor mounted at a high position on a streetlamp prevents that objects remain hidden behind other objects and therefore undetected.

The paradigm of infrastructure-based sensing can support other types of sensing beyond cameras. For example, radars mounted on infrastructure have the advantage of being "less intrusive" in cities where the population is resistant to video monitoring. LiDAR may also be employed to obtain a high-resolution snapshot of the environment. In Japan, RSUs equipped with LiDAR sensors have been deployed for commercial V2I services of ITS Connect in several regions of Japan such as Tokyo, Aichi, Osaka, Fukuoka, Miyagi, etc. [68]. The LiDAR sensor of an RSU installed at an intersection is used to detect pedestrians crossing the intersection and vehicles approaching the intersection in case that they do not support ITS Connect. After detecting them, the RSU notifies the surrounding vehicles supporting ITS Connect working on DSRC about the existence of the detected pedestrians and vehicles to avoid the potential collisions. It should also be noted that sensing from the RSU can be used to support the mmWave V2X operations, including providing information to make beam training more efficient [21][22]. Further research is still required to integrate these methods more deeply into the 5G NR and IEEE802.11ad/bd paradigm.

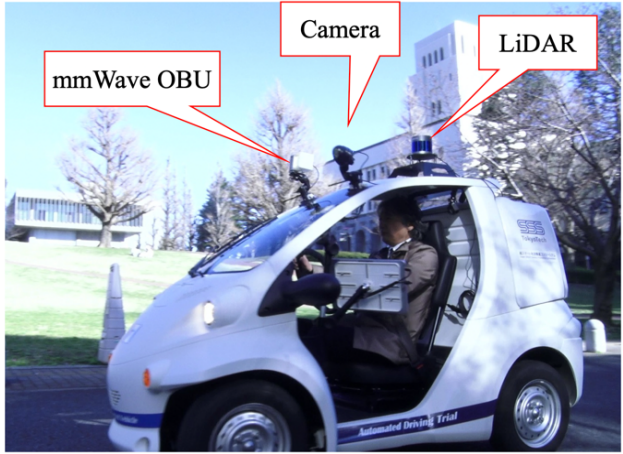

**Fig.22** Overview of research field of automated driving

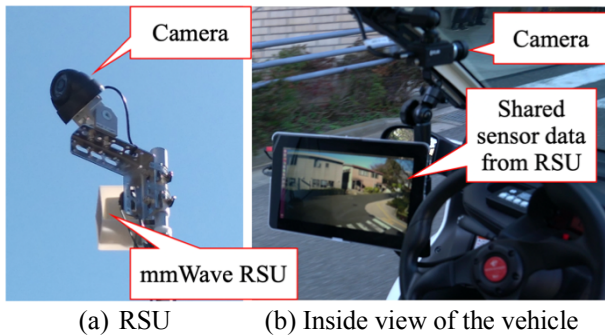

(a) RSU      (b) Inside view of the vehicle

**Fig. 23** Sensor data sharing between RSU and OBU

One of the research fields for automated driving using mmWave V2X is shown in Fig.22. The field is a part of SSS project and is placed at the Ōokayama campus of Tokyo Institute of Technology [70]. The field involves an RSU equipped with mmWave transmitter/receiver (TRX) at 60 GHz and an optical camera, while the OBU is equipped with a similar mmWave TRX, camera and LiDAR. The vehicle detects its surroundings by using the LiDAR system on the roof and comparing the position with the prior stored 3D map. The vehicle is equipped with "*Autoware*", an open source software operated on the Robot Operating System (ROS), and by utilizing all of these capabilities, the automated driving is carried out.

Pictures of the RSU and an inside view of the vehicle are shown in Fig. 23. The RSU was located at the corner of an intersection, where the camera observes the scene and transmits this information to the vehicle using mmWave. This allows the driver to detect hidden objects by monitoring the out-of-sight image on the monitor screen of the vehicle. Meanwhile, the video taken by the on-board camera is sent to the server via the RSU in order to provide the vehicle with an adequate guidance.

In the current configuration, the video taken by the optical camera on the RSU or OBU is not directly utilized when the vehicle is in automated driving mode. However, incorporation of such camera information combined with the LiDAR and the 3D map information allows the system to provide more accurate and safe driving.

## 5. Concluding Remarks

The commercialization of fully automated driving is expected around 2025. For safer automated driving, this paper described the importance of mmWave V2X for cooperative perception by exchanging sensor data with other vehicles and RSUs. It provided a survey on existing V2X standards without mmWave and explained three studies towards mmWave V2X, i.e. IEEE802.11ad/bd based V2X, extension of 5G NR, and prototypes of intelligent street with mmWave V2X.

This paper contributes to the next generation traffic systems for fully automated driving with infrastructure support, e.g. digital curve mirrors and digital traffic signals achieved by sensors and mmWave V2X. A revolutionized transportation system based on automated driving will change our daily life and contribute towards solving essential challenges such as exhausting commute experiences and transporting an aging society.

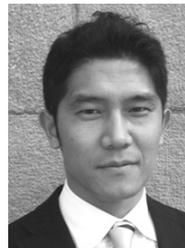

**Kei Sakaguchi** received the M.E. degree in Information Processing from Tokyo Institute Technology in 1998, and the Ph.D degree in Electrical & Electronics Engineering from Tokyo Institute Technology in 2006. Currently, he is working at Tokyo Institute of Technology in Japan as a Dean in Tokyo Tech Academy for Super Smart Society and as a Professor in School of Engineering. At the same time, he is working for oRo Co.,Ltd. in Japan as an outside director. He received the Outstanding Paper Awards from SDR Forum and IEICE in 2004 and 2005 respectively, and three Best Paper Awards from IEICE communication society in 2012, 2013, and 2015. He also received the Tutorial Paper Award from IEICE communication society in 2006. His current research interests are in 5G cellular networks, millimeter-wave communications, wireless energy transmission, V2X for automated driving, and super smart society. He is a fellow of IEICE, and a member of IEEE.

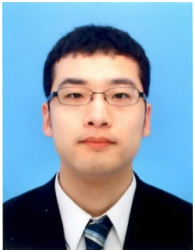

**Ryuichi Fukatsu** received the B.E. and M.E. degrees in electrical and electronic engineering from Tokyo Institute of Technology, Japan, in 2017 and 2019 respectively. He is currently a Ph.D. student at Tokyo Institute of Technology. His research interests are V2V and automated driving.

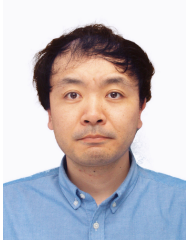

**Tao Yu** received the M.E. degree in signal and information processing from Communication University of China in 2010, and the Dr.Eng. degree in electrical and electronic engineering in 2017 from Tokyo Institute of Technology, where he is working as a postdoctoral researcher. His current research interests are UAV communication, 5G, wireless sensor networks, wireless energy transmission. He is a member of IEICE and IEEE.

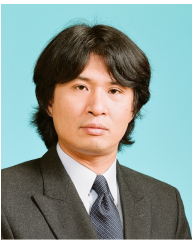

**Eisuke Fukuda** graduated from the faculty of electrical engineering of Tohoku University in 1979 and joined Fujitsu Laboratories Ltd., where has a 30-years-long R&D experience on wireless communication systems. Since 1993, he had been engaged in research work for the 3rd generation mobile communications system such as W-CDMA or IMT-2000 as well as the 4th generation such as LTE and LTE-Advanced. In addition, he had actively been involved in global standardization as Vice Chairman of TSG-RAN of 3GPP from 2001 to 2005 and a regular member of Japan's delegates of ITU-R SG8 WP8F. Since 2019, he has been with Tokyo Institute of Technology as Specially Appointed Professor. He has a doctorate from the University of Tokyo. He is a fellow of the Institute of Electronics, Information, and Communication Engineers (IEICE) of Japan and a member of the Institute of Electrical and Electronics Engineers (IEEE). His current interest includes the 5G systems and the related various services.

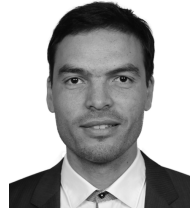

**Kim Mahler** is a Smart Cities Postdoc at the New York University. His research interests involve the development of viable 5G applications, drone/vehicular communications, localization using 5G millimeter wave and user-centric innovation developments. He received the M.Sc. degree with honors and the Dr.-Ing. degree (magna cum laude) from the Technical University of Berlin, Germany, in 2010 and 2016. He also received the M.A. degree from the Berlin University of Arts/University of St. Gallen in 2014. From 2010 until 2019, he was with the Wireless Communications and Networks department at Fraunhofer Heinrich Hertz Institute HHI in Berlin, Germany.

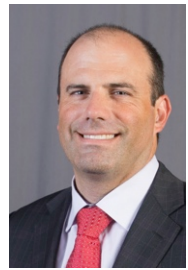

**Robert W. Heath Jr.** is a Cockrell Family Regents Chair in Engineering in the Department of ECE at The University of Texas at Austin, and Director of UT SAVES. He authored "Introduction to Wireless Digital Communication" (Prentice Hall in 2017) and co-authored "Millimeter Wave Wireless Communications" (Prentice Hall in 2014) and Foundations of MIMO Communication (Cambridge in 2018). He is a licensed Amateur Radio Operator, a registered Professional Engineer in Texas, a Private Pilot, recipient of the 2017 EURASIP Technical Achievement Award, the 2019 IEEE Kiyo Tomiyasu Award, is a member of EURASIP and IEICE, a Fellow of the National Academy of Inventors, and is a Fellow of the IEEE.

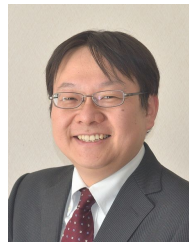

**Takeo Fujii** received the B.E., M.E. and Ph.D. degrees in electrical engineering from Keio University, Yokohama, Japan, in 1997, 1999 and 2002 respectively. From 2002 to 2006, he was an assistant professor in the Department of Electrical and Electronic Engineering, Tokyo University of Agriculture and Technology. Since 2006, he has been working at The University of Electro-Communications. Currently, he is a professor in Advanced Wireless and Communication Research Center, The University of Electro-Communications. His current research interests are in spectrum sharing, spectrum database and wireless distributed networks. He received Best Paper Award in IEEE VTC 1999-Fall, 2001 Ericsson Young Scientist Award, Young Researcher's Award from the IEICE, Best Paper Award in IEEE CCNC 2013, and IEICE Communication Society Best Paper Award in 2016. He is a member of IEEE.

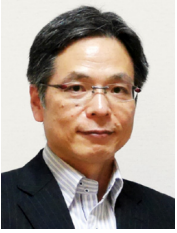

**Kazuaki Takahashi** received the B.E. and M.E. degrees in electrical and computer engineering, and the Ph.D. degree in electrical engineering from Yokohama National University, Japan, in 1986, 1988 and 2006. In1988, he joined the Tokyo Research Laboratory, Matsushita Electric Industrial Co. Ltd., Kawasaki Japan, where he was engaged in research and development of monolithic microwave ICs, millimeter wave ICs based on Si and GaAs for mobile communication equipment. Also, he was engaged in Si micromachined packaging technology for millimeter wave system integration. His current research interests include the development of short-range multi-gigabit wireless system and high-resolution radar system in millimeter wave. He is currently a Chief Engineer with Communication Core Devices Development Center, Panasonic Corporation., Yokohama Japan. He a member of the IEEE.

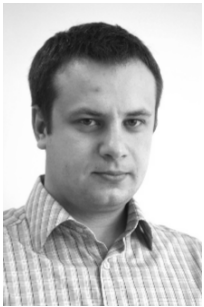

**Alexey KHORYAEV** is a principal engineer at Intel Next Generation and Standards Group. He received his M.S. degree in radio physics from Nizhny Novgorod State University, Russia in 2002. He has been working on 3GPP LTE standards since 2011. His research interests are in the areas of physical layer design and upper layer protocols for wireless communication systems with focus on sidelink and V2X communication systems during the last years.

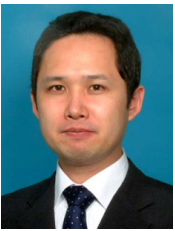

**Satoshi Nagata** received his B.E. and M.E. degrees from Tokyo Institute of Technology, Tokyo, Japan, in 2001 and 2003, respectively. In 2003, he joined NTT DOCOMO, INC. He worked for the research and development for wireless access technologies for LTE, LTE-Advanced, and 5G. He is currently a manager working for 5G/6G and 3GPP standardization. He had contributed to 3GPP over ten years, and contributed 3GPP TSG-RAN WG1 as a vice chairman from November 2011 to August 2013, and as a chairman from August 2013 to August 2017. He is currently a vice chairman of 3GPP TSG-RAN since March 2017. He was a recipient of the 2003 active research award in radio communication systems from IEICE in 2003.

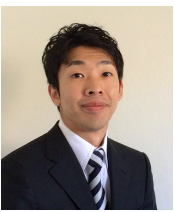

**Takayuki Shimizu** received the B.E., M.E., and Ph.D. degrees from Doshisha University, Kyoto, Japan, in 2007, 2009, and 2012, respectively. From 2009 to 2010, he was a visiting researcher at Stanford University, CA, USA. From 2012 to 2019, he worked at TOYOTA InfoTechnology Center, U.S.A., Inc. Currently, he is a Principal Researcher at Toyota Motor North America, Inc., where he works on the research and standardization of wireless vehicular communications. His current research interests include millimetre-wave vehicular communications, vehicular communications for automated driving, and LTE/5G for vehicular applications. He is a co-author of the NOW monograph entitled "Millimeter Wave Vehicular Communications: A Survey" published by NOW Publishers in 2016. He is a 3GPP RAN delegate for V2X standardization and a member of multiple SAE Technical Committees for V2X. He received the 25th TELECOM System Technology Award for Student from the Telecommunications Advancement Foundation in 2010. He is a member of the IEEE.